\documentclass[10pt,twocolumn,english,aps,prl,superscriptaddress,showpacs]{revtex4}
\usepackage{graphicx,psfrag,amsmath,amssymb,float}
\usepackage{graphicx}
\usepackage{epstopdf,times}

\usepackage{graphicx,natbib,ulem}
\usepackage[colorlinks,bookmarks=false,citecolor=blue,linkcolor=red,urlcolor=blue]{hyperref}

\usepackage{color}
\newcommand{\be}{\begin{equation}}
\newcommand{\ee}{\end{equation}}
\newcommand{\bea}{\begin{eqnarray}}
\newcommand{\eea}{\end{eqnarray}}

\newcommand{\bc}{\begin{center}}
\newcommand{\ec}{\end{center}}
\begin{document}
\title{Theory of the field-induced BEC in the frustrated spin-$\frac{1}{2}$ dimer compound BaCuSi$_2$O$_6$}
\author{Nicolas Laflorencie}
\affiliation{Laboratoire de Physique des Solides, 
Universit\'e Paris-Sud, UMR-8502 CNRS, 91405 Orsay, France}
\author{Fr\'ed\'eric Mila}
\affiliation{Institute of Theoretical Physics, \'Ecole Polytechnique F\'ed\'erale de Lausanne, CH-1015 Lausanne, Switzerland}
%
\begin{abstract}
Building on recent neutron and NMR experiments, we investigate the field-induced exotic 
criticality observed in the frustrated spin-$\frac{1}{2}$ dimer compound BaCuSi$_2$O$_6$ using a frustrated model with two types of bilayers. A semiclassical treatment
of the effective hard-core boson model shows that perfect interlayer 
frustration leads to a 2D-like critical exponent
$\phi=1$ without logarithmic corrections and to a 3D low temperature phase with
different but non vanishing triplet populations in both types of bilayers. These further suggest
a simple phenomenology in terms of a field-dependent transverse coupling in the context of which we reproduce the entire field-temperature phase diagram with Quantum Monte Carlo simulations.
\end{abstract}
\maketitle
\noindent{\it{Introduction---}}~Field-induced Bose-Einstein condensation (BEC) of the excitations in antiferromagnetic (AF) spin systems is a fascinating phenomenon. Theoretically predicted some years ago~\cite{Affleck91,Giamarchi99}, it was first observed experimentally in the spin-$\frac{1}{2}$ dimer compound TlCuCl$_3$~\cite{Oosawa99,Nikuni00,Ruegg03}. Recently, a very intense activity has emerged~\cite{Giamarchi08}, showing a large and rich variety of field-induced bosonic phases in such quantum materials. Interestingly, a {\it quantitative} theoretical description is often possible through effective theories written in term of hard-core bosons. Such theories have been quite successful to explain various non-trivial features such as magnetization plateaus in spin ladders~\cite{Totsuka98-Mila98,Ladders} and frustrated lattices like Sr$_2$Cu(BO$_3$)$_2$~\cite{Takigawa02,Dorier-Sylvain}, field-induced Luttinger liquid behavior in quasi-1D systems~\cite{LL}, or even a more exotic phase, analog to a supersolid, where a uniform triplet condensate coexists with a spontaneous spatial modulation of the triplets~\cite{Ng06-Laflorencie07}. 

\begin{figure}
\begin{center}
\includegraphics[width=0.9\columnwidth,clip]{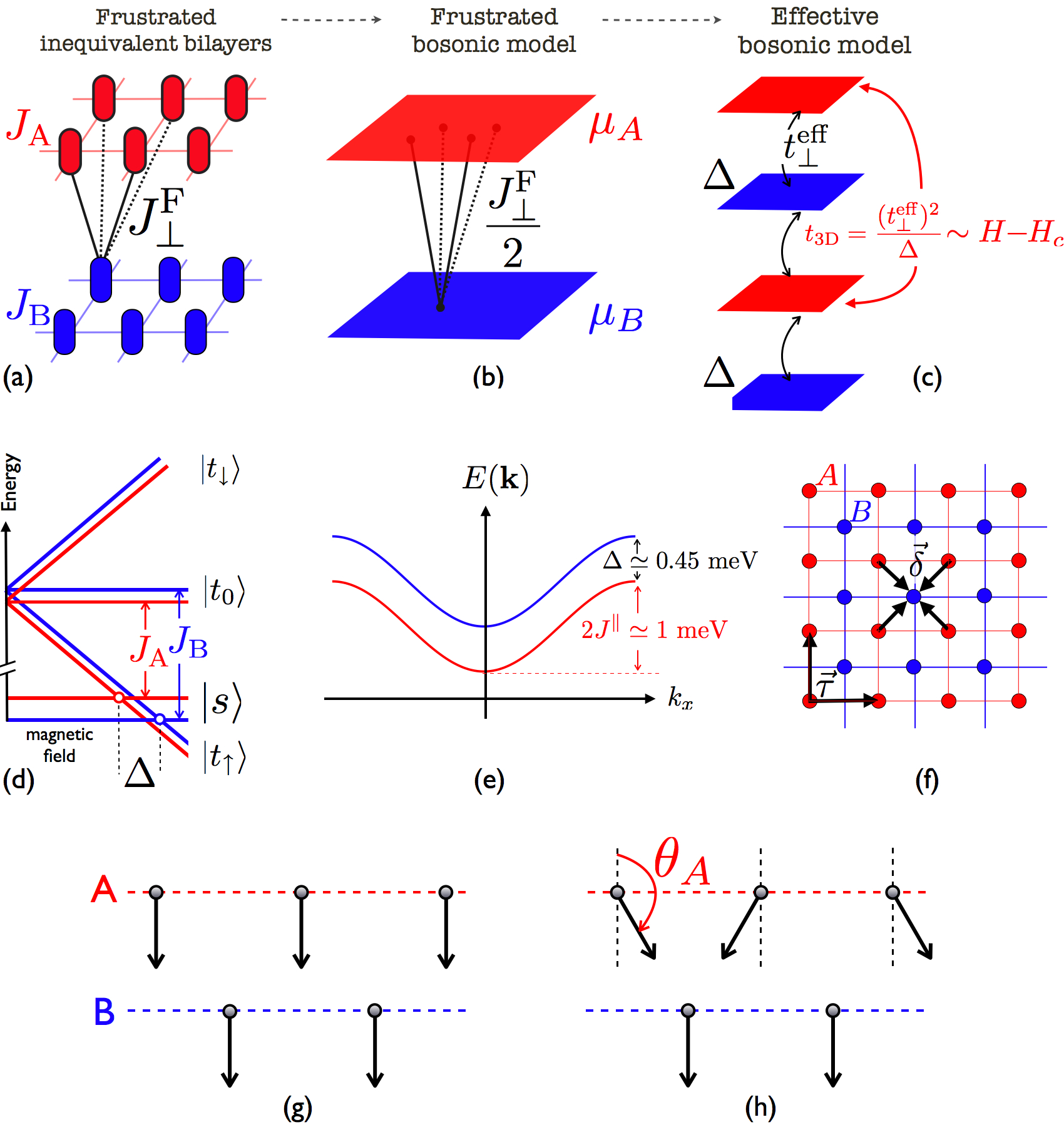}
\end{center}
\vskip -0.5cm
\caption{Schematic models for BaCuSi$_2$O$_6$: (a) Frustrated bilayer array with two types of dimers; (b) and (c) Effective bosonic models in the field-induced critical regime: (b) Frustrated hard-core bosons model with different chemical potential; (c) Non-frustrated hard-core boson model with a field-dependent effective transverse hopping (see text); (d) The 4 energy states of isolated dimers A and B versus the external magnetic field;
(e) Triplet dispersions in isolated bilayers  $E_{\rm A}=J^{\parallel}\left(\cos k_x+\cos k_y\right)$ and $E_{\rm B}=E_{\rm A}+\Delta$ induced by the inter-dimer couplings $J^{\parallel}$;
(f)View of the frustrated lattice (a) from above with the unit vectors $\vec\tau$ and $\vec\delta$. Bottom: Angle representation of the classical GS. (g) $\mu<\mu_c$: both A and B layers are empty. (h) $\mu_c<\mu<\mu_c+\Delta$: A layers start to accomodate bosons while B layers are still gapped.}
\label{fig:Pict}
\end{figure}

Another remarkable behavior has been reported 
in BaCuSi$_2$O$_6$, a 3D stack of spin-$\frac{1}{2}$ bilayers with frustrated interlayer coupling known as Han purple. 
This compound undergoes a field-induced AF ordering described as a triplet BEC~\cite{Jaime2004}, but the field dependence of the critical 
temperature above the quantum critical point (QCP) at $H_c\simeq 23.2$ T
has an anomalous exponent:~\cite{Sebastian2006} 
\be
T_{\rm BEC}(H)\sim (H-H_c)^{\phi}
\label{eq:Tc}
\ee
with $\phi\simeq1$. Since the BEC has an exponent $\phi=2/D$,
this behavior is typical of a 2D system. This lead Sebastian {\it et al.}~\cite{Sebastian2006} to interpret 
their result as a frustration driven ``dimensional reduction'' of the interlayer coupling $J_{\perp}^{\rm F}$ based on the observation that the triplet transverse dispersion
$
\sim J_{\perp}^{\rm F}\cos\left({k_x}/{2}\right)\cos\left({k_y}/{2}\right)\cos k_z
$
is flat at the AF condensate wave vector $(\pi,\pi)$, so that there is no 3D coherence of the condensate. 
This was later confirmed by inelastic  neutron scattering experiments~\cite{Ruegg2007}, which indeed
reported an almost flat 3D dispersion at $(\pi,\pi)$, but which also revealed a richer structure with inequivalent bilayers
of at least two types: $J_{\rm A}=4.27(1)$ meV, $J_{\rm B}=4.72(1)$ meV~\cite{Ruegg2007}, as depicted in Fig.~\ref{fig:Pict}(a). The presence of at least two types of layers has been confirmed by high field NMR experiments which have identified a larger triplet population in A-layers than in B ones.~\cite{Kramer2007}. These NMR results point however to the 3D nature of the ordered
phase at low temperature since B-layers have a small but non zero population above $H_c$.

Taken together, these experimental results~\cite{Sebastian2006,Ruegg2007,Kramer2007} lead to a fundamental puzzle: How can a 3D bosonic population lead to a 2D critical exponent for the BEC transition line? And more generally, what is the exact role played by the frustration? Two theories have been put forward, with somehow incompatible conclusions. Batista {\it et al.}~\cite{Batista2007} have investigated a model with equivalent layers. As ancitipated in Ref.~\cite{Sebastian2006}, they found that, although the low temperature phase is 3D, the critical temperature does not have the 3D exponent $\phi=2/3$. They obtained $\phi=1$, but with logarithmic corrections. This conclusion holds provided the interlayer coupling is fully frustrated. By contrast,
R\"osch and Vojta~\cite{Rosch2007} have investigated a model with different bilayers and came to the
conclusion that the interlayer coupling cannot be fully frustrated since this would be inconsistent
with the NMR observation of a non zero triplet population in B-layers just above $H_c$. 

In this Letter, we show that all these experimental data can in fact be reconciled in the context of a
model including the two types of bilayers {\it and} perfect frustration. In particular, we show that the 
presence of two types of bilayers
lead to $T_{\rm BEC}(H)\sim (H-H_c)$ {\it without logarithmic corrections}, while the B-layers
get a triplet population immediately above $H_c$. However, this population only grows
as $\rho_{\rm{B}}\propto (H-H_c)^2$, i.e. much more slowly than that in the A-layers 
($\rho_{\rm{A}}\propto H-H_c$). 
Starting form a 3D frustrated spin bilayer model with two types of bilayers, these conclusions 
have been reached by investigating the spin-wave (SW) corrections around the classical 
ground-state (GS) of an effective frustrated bosonic model for the triplets with an
energy barrier between the A and B planes of $\Delta\simeq 0.45$ meV~\cite{Ruegg2007}.
These results can be reinterpreted as a field-dependent effective quantum tunnelling in the transverse direction $t_{\rm 3D}\sim (H-H_c)$ for the triplets.  This naturally implies 
a 3D critical BEC temperature with an effective 2D-like exponent
$
\phi=\frac{2}{D}+\frac{1}{3}
$
and has allowed us to reproduce the entire experimental $T-H$ phase boundary using large scale quantum Monte Carlo (QMC) simulations of a 3D effective bosonic model with a field-dependent tunneling.

\noindent{\it{Effective model---}}~The theoretical analysis starts from a magnetic Hamiltonian describing a 3D array of frustrated bilayers with two types of dimers [Fig.~\ref{fig:Pict}(a)] that we can decompose into three parts: 
${\cal H}^{\rm mag}={\cal{H}}^{\rm mag}_{A}+{\cal{H}}^{\rm mag}_{B}+{\cal{H}}^{\rm mag}_{\perp}$.
The bilayers A and B are described by the Hamiltonians
\bea
{\cal{H}}^{\rm mag}_{{\rm{A(B)}}}&=&\sum_{{\vec{r}}\in {\rm{A(B)}}}\Bigl[J_{{\rm{A(B)}}}{\bf{S}}_{{\vec{r}},1}\cdot{\bf{S}}_{{\vec{r}},2}-H(S_{{\vec{r}},1}^{z}+S_{{\vec{r}},2}^{z})\nonumber\\
&+&\sum_{l=1,2}
\sum_{{\vec \tau}} J^{\parallel}_{{\rm{A(B)}}} {\bf{S}}_{{\vec{r}},l}\cdot{\bf{S}}_{{\vec{r}}+{\vec{\tau}},l}\Bigr],
\label{eq:H1}
\eea
with $J_{\rm A}\simeq 4.27$ meV, $J_{\rm B}\simeq 4.72$ meV, and $J^{\parallel}_{\rm A}\simeq J^{\parallel}_{\rm B}\simeq 0.5$ meV for BaCuSi$_2$O$_6$~\cite{Ruegg2007}.
The frustrated inter-bilayer part is governed by
\be
{\cal{H}}^{\rm mag}_{\perp}=\sum_{{\vec{r}}\in {\rm{A}}}\sum_{{\vec{\delta}}}
J_{\perp}^{\rm F}\Bigr[{\bf{S}}_{{\vec{r}},2}\cdot{\bf{S}}_{{\vec{r}}+{\vec{\delta}},1}+{\bf{S}}_{{\vec{r}},1}\cdot{\bf{S}}_{{\vec{r}}-{\vec{\delta}},2}\Bigr].
\label{eq:H2}
\ee
Note that the space coordinates
are labelled by 
$\vec r,l$ where $l=1,2$ are the layer indices within each bilayers, and elementary displacements $\vec\tau$ and $\vec\delta$ are defined in Fig.~\ref{fig:Pict}(f).
Since we consider the strong coupling limit ($J_{\rm A,B}\gg J^{\parallel}$), each bilayer can be described by a low energy theory where only the two lowest states of a dimer (the singlet $|s\rangle=\left(|\hskip-0.1cm \uparrow\downarrow\rangle-|\hskip-0.1cm\downarrow\uparrow\rangle\right)/\sqrt{2}$ and the polarized triplet $|t_{\uparrow}\rangle=|\hskip-0.1cm\uparrow\uparrow\rangle$ shown in Fig.~\ref{fig:Pict}(d)) are retained, each dimer being replaced by a single site with hard-core boson degrees of freedom ($|0\rangle\equiv |s\rangle$ and $|1\rangle\equiv |t_{\uparrow}\rangle$)~\cite{Totsuka98-Mila98,Giamarchi99}. Each AF bilayer is then described by interacting hard-core bosons on a 2D square lattice, and the 3D effective frustrated bosonic Hamiltonian [Fig.~\ref{fig:Pict}(b)] reads
${\cal H}^{\rm bos}={\cal{H}}_{\rm A}+{\cal{H}}_{\rm B}+{\cal{H}}_{\perp}$, with
\be
{{\cal H}}_{\rm A}=\sum_{\vec{r},{\vec{\tau}}}
\frac{J^{\parallel}}{2}(a^{\dagger}_{\vec r}a^{\vphantom{\dagger}}_{\vec r+\vec \tau}+{\rm{h.c.}}
+n^{\rm A}_{\vec r}n^{\rm A}_{\vec r+\vec\tau})-\sum_{\vec{r}}\mu_{\rm A}n^{\rm A}_{\vec r}
\label{eq:A}
\ee
and
\be
{{\cal H}}_{\rm B}=\sum_{\substack{\vec{r},{\vec{\tau}}}}
\frac{J^{\parallel}}{2}(b^{\dagger}_{\vec r}b^{\vphantom{\dagger}}_{\vec r+\vec \tau}+{\rm{h.c.}}
+n^{\rm B}_{\vec r}n^{\rm B}_{\vec r+\vec\tau})-\sum_{\vec{r}}\mu_{\rm B}n^{\rm B}_{\vec r},
\label{eq:B}
\ee
where $\mu_A=J_A-H$ and $\mu_B=J_B-H=\mu_A+\Delta$. The frustrating inter-layer part
mixes bosons $a$ and $b$:
\be
{\cal{H}}_{\perp}=\sum_{\vec r\in {\rm A},\vec\delta}\frac{J_{\perp}^{\rm F}}{2}(a^{\dagger}_{\vec r}b_{\vec r\pm\vec \delta}+
b^{\dagger}_{\vec r}a_{\vec r\pm\vec \delta}+n^{\rm A}_{\vec r}n^{\rm B}_{\vec r\pm\vec\delta}).
\label{eq:perp}
\ee

\noindent{\it{Mean-Field analysis---}}~We first analyse the problem using a mean-field (MF) Hartree-Fock approach, searching for a GS wave function of the form $\Psi_0=\psi_{\rm A}\otimes\psi_{\rm B}$ with
$|\psi_{\rm A}\rangle=\prod_{j\in{\rm A}}\left(\sin{\frac{\theta_{\rm A}}{2}}|0\rangle_{j}+{\rm e}^{i\varphi_j}\cos\frac{\theta_{\rm A}}{2}|1\rangle_{j}\right)$.
When decoupled, each layer displays successive phases upon changing the chemical potential. If $\mu<-2J^{\parallel}$ ($>2J^{\parallel}$) the bosonic system is empty (full): $\theta=\pi$ ($\theta=0$) and $|\psi\rangle=\prod_j|0\rangle_j$ ($|\psi\rangle=\prod_j|1\rangle_j$) whereas for $-2J^{\parallel}\le \mu\le 2J^{\parallel}$ the system is compressible and 
the bosons can form a $T=0$ Bose-Einstein condensate with $0<\theta<\pi$ and a fixed (pinned) phase $\varphi$ (finite superfluid stiffness). Interestingly, the layers A and B start to accommodate bosons for different values of $\mu$, meaning different values of the magnetic field $H$,
resulting in the distinct triplet branches shown in Fig.~\ref{fig:Pict}(e). When frustration [Eq.~(\ref{eq:perp})] is taken into account, the gap remains open at the MF level because frustration leads to 
a cancellation of the hopping processes between A and B and only allows for non-zero diagonal interactions which just renormalize the gap $\Delta=\left({J_{\rm B}-J_{\rm A}}\right)/\left({1-\frac{J_{\perp}^{\rm{F}}}{3{J^{\parallel}}}}\right)$.
Therefore, at the MF level the triplet densities $\rho_{{\rm{A(B)}}}=(1+\cos  {\theta_{{\rm{A(B)}}}})/2$ appear at different critical fields with a quite sizable gap $\Delta \sim 3.5$ T. However, 
as already discussed by R\"osch and Vojta~\cite{Rosch2007}, the existence of a second critical field where triplets would start to populate the B planes is ruled out by NMR results~\cite{Kramer2007} which clearly show a unique critical field with a finite, while reduced, triplet population on the B layers.
Therefore, one needs to go beyond MF and consider the effect of quantum fluctuations around the classical GS in order to understand how the triplets populate the B sites, and how they propagate along the 3D direction.

\begin{figure}
\begin{center}
\includegraphics[width=\columnwidth,clip]{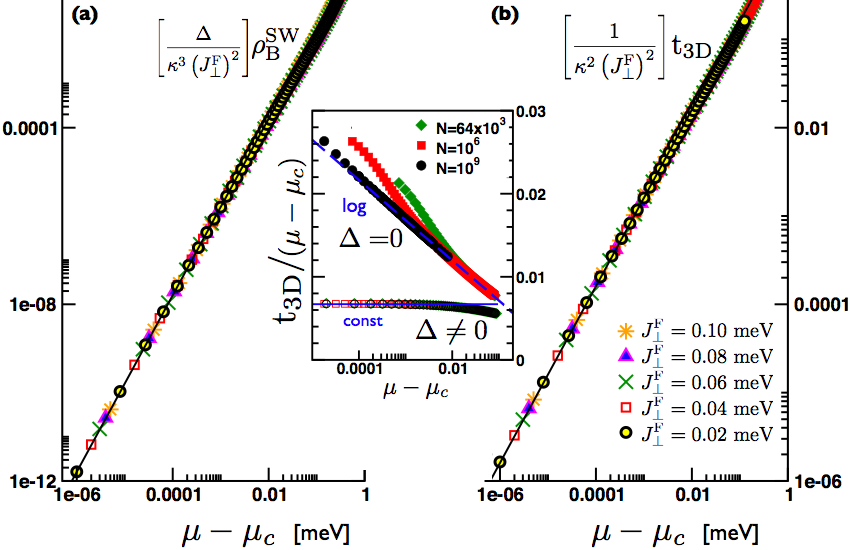}
\end{center}
\vskip -0.5cm
\caption{Scaling plots from SW calculations with realistic parameters $J_{\parallel}=0.5$ meV, $\Delta=0.45$ meV and various values of the frustrating couplings $J_{\perp}^{\rm F}$ indicated on the graph. 
(a) Quadratic behavior of $\rho_{\rm B}^{\rm SW}$ Eq.~(\ref{eq:rhob}); (b) Linear behavior for the 3D tunnelling t$_{\rm 3D}$ Eq.~(\ref{eq:t3d}) deduced from the transverse stiffness $\rho_{\perp}^{\rm sf}$. Black lines are fits using Eqs.~(\ref{eq:rhob},\ref{eq:t3d}) with ${\rm{const}}=1.6$. Inset: Comparison between equivalent layers ($\Delta=0$, closed symbols) and non-equivalent layers ($\Delta=0.5$ meV, open symbols). The absence of finite size effects and the linear behavior $t_{\rm 3D}\propto \mu-\mu_c$ in the latter case contrast with the strong logarithmic corrections of the former case.}
\label{fig:scaling}
\end{figure}

\noindent{\it{Linear spin-waves---}}~
We replace a hard-core boson by a spin-$1/2$ with classical coordinates ${\vec{S}}=1/2\left(\cos\varphi\sin\theta,\sin\varphi\sin\theta,\cos\theta\right)$, leading to the vector representation depicted in Fig.~\ref{fig:Pict}(g-h). 
We develop the SW calculation around these classical vectors, including the finite external field~\cite{SWexample,TBP}. After a few steps, 
%
we compute the SW-corrected densities $\rho^{\rm SW}_{\rm A}$ and $\rho_{\rm B}^{\rm SW}$ which both appear for $\mu>\mu_c$.
While $\rho_{\rm A}^{\rm SW}$ remains linear with $\mu-\mu_c$, SW fluctuations close the "classical gap" $\Delta$ and induce a small triplet population in B layers as soon as $\mu>\mu_c$. This bosonic occupation being classically forbidden in the region (b) $\mu<\mu_c+\Delta$, it is purely induced by quantum fluctuations. More precisely, an unexpected quadratic dependence on the external field is found, with the following dependence 
\be
\rho_{\rm B}^{\rm SW}\simeq{\rm{const}}\times\frac{\left(J_{\perp}^{\rm F}\right)^2}{\Delta} \kappa^3 (\mu-\mu_c)^2,
\label{eq:rhob}
\ee
$\kappa=(6J^{\parallel})^{-1}$ being the compressibility of an isolated layer, and ${\rm const}$ is ${\cal{O}}(1)$ (see Fig.~\ref{fig:scaling}).
SW results are shown in Fig.~\ref{fig:scaling}(a) with realistic parameters for BaCuSi$_2$O$_6$ ($J_{\parallel}=0.5$ meV and $\Delta=0.45$ meV) and various values of the frustrating coupling $J_{\perp}^{\rm F}$. 
Such a quadratic dependence, which
is compatible with current NMR data~\cite{Kramer2007,berthier}, has a dramatic effect on the 3D criticality of the system.
Indeed, considering the {\it non-frustrated} version of this problem: a system having a similar energy barrier $\Delta$ between adjacent planes but with non-frustrating inter-layer tunnelling t$_{\perp}$, one obtains a unique QCP with a B occupation growing linearly like $\rho_{\rm B}\simeq t_{\perp}^{2}\kappa(\mu-\mu_c)/{\Delta^2}$. This implies that the quadratic population in the frustrated case Eq.~(\ref{eq:rhob}) results from an effective field-dependent inter-layer tunnelling amplitude $t_{\perp}^{\rm eff}\sim \sqrt{\mu-\mu_c}$. Therefore the 3D coherence, governed by the $2^{\rm nd}$ order hopping process [Fig.~\ref{fig:Pict}(c)], is controlled by the energy scale
\be
t_{3{\rm D}}=\frac{\left(t_{\perp}^{\rm eff}\right)^2}{\Delta}\simeq{\rm{const}}\times \left(J_{\perp}^{\rm F}\right)^2 \kappa^2 (\mu-\mu_c).
\label{eq:t3d}
\ee

As an alternative check for such a field-dependence for the effective 3D tunnelling amplitude, we introduce a Peierls phase twist $\varphi_{\perp}$ in the transverse direction and compute the transverse superfluid stiffness 
$\rho_{\perp}^{\rm sf}={\partial^2 \langle {\cal{H}}(\varphi_{\perp})\rangle}/{\partial \varphi_{\perp}^2}\bigl|_{\varphi_{\perp}=0}$. A finite stiffness signals the appearance of 3D phase coherence, hence of an effective hopping $t_{\rm 3D}$ with
$\rho_{\perp}^{\rm sf}\simeq t_{\rm 3D}\sin^2\theta_A$.
SW results for $t_{\rm 3D}\simeq\rho_{\perp}^{\rm sf}/[\rho_{\rm A}(1-\rho_{\rm A})]$ are shown in Fig.~\ref{fig:scaling}(b) for realistic parameters. The agreement with the linear behavior Eq.~(\ref{eq:t3d}) is excellent. On the other hand, {\it{equivalent}} layers ($\Delta=0$) do not lead to a simple linear 3D hopping but to quite strong logarithmic corrections (inset of Fig.~\ref{fig:scaling}).

\noindent{\it{Consequences for the BEC critical temperature---}}~The kinetic energy of a diluted {\it{isotropic}} Bose gas is $\epsilon({\vec q})=\hbar^2{\vec q}^2/(2m^*)$. If D$>2$ there is a finite-$T$ BEC when the De Broglie length becomes greater than the inter-particle distance $d\sim \rho^{-\frac{1}{D}}$, at
$
T_{\rm c}\propto \frac{\hbar^2}{2m^*}\rho^{\frac{2}{D}}
$.
In our specific case, close to the critical field, the kinetic energy is {\it{anisotropic}}:
{\protect{$\epsilon({\vec{q}})=t\left(q_{x}^{2}+q_{y}^{2}\right)+t_{\rm 3D}q_{z}^{2}$}}. The linear $t_{\rm 3D}(\mu)$ leads to an effective mass $m^*\sim (\mu-\mu_c)^{-\frac{1}{D}}$, thus implying a modified 3D BEC temperature
\be
T_{\rm c}\propto \frac{(\mu-\mu_c)^{\frac{2}{D}}}{m^*}\propto \mu-\mu_c,
\label{eq:lineartc}
\ee
with an effective exponent $\phi=\frac{2}{D}+\frac{1}{D}=1$, as experimentally measured~\cite{Sebastian2006}. It is important to note that with only one type of layer ($\Delta=0$), the logarithmic corrections discussed above for t$_{\rm 3D}$ will show up in $T_c(H)$, as also found in Refs.~\cite{Batista2007} using a different theoretical approach.
\begin{figure}
\begin{center}
\includegraphics[width=\columnwidth,clip]{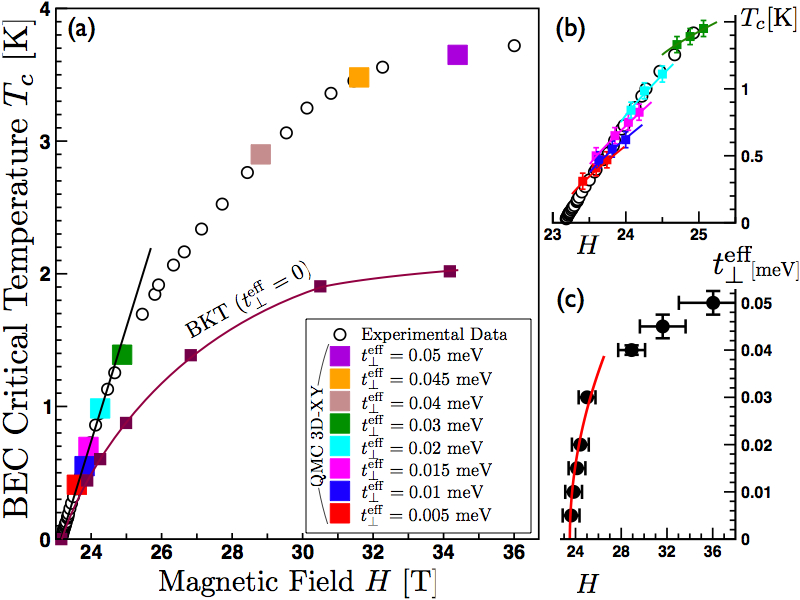}
\end{center}
\vskip -0.5cm
\caption{(a) Temperature-magnetic field phase boundary for BaCuSi$_2$O$_6$. Experimental data~\cite{Sebastian2006} (open circles) are confronted to QMC results (colored squares) obtained with a field-dependent transverse hopping $t_{\perp}^{\rm eff}$. The solid line is a guide to the eyes for the linear part. The 2D-BKT transition line of a single decoupled bilayer is also shown for comparison. (b) Zoom over the critical linear regime where more QMC data points are shown. (c) Effective hopping $t_{\perp}^{\rm eff}$ used  to fit the experimental phase boundary. The red curve is the theoretical $t^{\rm eff}_{\perp}$ using Eq.~(\ref{eq:t3d}), the best fit being obtained for $H_c\simeq 23.5(3)$ and $J_{\perp}^{\rm F}\simeq 0.15(4)$ meV.}
\label{fig:QMC}
\end{figure}

\noindent{\it{Quantum Monte Carlo---}}~A direct test of these conclusions on the model of Eqs.~(\ref{eq:H1},\ref{eq:H2}) with QMC is impossible because of the severe minus sign problem induced by frustration. In order to test the internal logic of the previous arguments regarding the linear field-dependence of the critical temperature Eq.~(\ref{eq:lineartc}), we performed QMC simulations on the {\it non-frustrated} effective bosonic model [Fig.~\ref{fig:Pict}(c)] ${\cal{H}}_{\rm eff}={\cal{H}}_{\rm A}+{\cal{H}}_{\rm B}+{\cal{H}}_{\perp}^{'}$, where ${\cal{H}}_{\rm A}$ and ${\cal{H}}_{\rm B}$ are given by Eqs.~(\ref{eq:A}) and (\ref{eq:B}), and ${\cal{H}}_{\perp}^{'}$ is the non-frustrated version of Eq.~({\ref{eq:perp}). Taking realistic parameters for BaCuSi$_2$O$_6$: $J^{\parallel}=0.5$ meV and $\Delta=0.45$ meV, the transverse hopping $t^{\rm eff}_{\perp}$ is adjusted such that the entire experimental phase boundary is reproduced. The inverse mapping bosons~$\to$~dimers is done such that the $T=0$ critical field $H_c$ takes the experimental value $\simeq 23.2$T for $t_{\perp}^{\rm eff}=0$. We achieved large-scale QMC calculations on systems of size ${\rm{L}}\times {\rm{L}}\times {\rm{L}}/2$ ($8\le {\rm{L}}\le 48$) with $0\le t_{\perp}^{\rm eff}\le 0.05$ meV. The BEC transition was revealed by stiffness crossings of $\rho_s(T_{c})\times L$, as expected for a 3D-XY transition. Shown in Fig.~\ref{fig:QMC}(a-b), the experimental data points are reproduced for the entire BEC critical dome using a continuously varying  transverse coupling $t_{\perp}^{\rm eff}$ plotted in Fig~\ref{fig:QMC}(c). A fit to Eq.~(\ref{eq:t3d}) leads to a rough estimate of the frustrating coupling $J_{\perp}^{\rm F}\simeq 0.15(4)$ meV.
As a comparison, the 2D BKT transition for a single layer with $J^{\parallel}=0.5$ meV is also displayed in Fig.~\ref{fig:QMC}(a). 
The curvature is clearly incompatible with the experimental data, thus demonstrating that the linear behavior of $T_c$ cannot be simply attributed to a 2D BKT transition.

\noindent{\it{Summary and experimental consequences---}}~ 
We have shown that the combined effect of two types of bilayers and of inter-layer frustration
in BaCuSi$_2$O$_6$ solves all puzzles regarding the physics close to the critical field $H_c$. It
explains the linear behavior of $T_c$ with $H-H_c$ reported in Ref.~\cite{Sebastian2006},
removing the logarithmic corrections of the single-bilayer model investigated in Ref.~\cite{Batista2007}, and it accounts for a small but non-zero triplet population in the B-layers.
These results also point to a phenomenological description in terms of a field-dependent effective transverse
hopping $t_{\rm 3D}\sim H-H_c$ which reconciles the concept of ``dimensional reduction'' at the QCP ($t_{\rm 3D}=0$ at $H_c$) with the 3D coherence of the low-$T$ phase above $H_c$.

This model also makes a number of predictions of experimental relevance. Firstly, 
the triplet population in the B-layers is quadratic with $H-H_c$. This behavior
is compatible with current NMR data~\cite{berthier}, but more precise NMR measurements
are needed to actually check this power law. Secondly, our estimate $J_{\perp}^{\rm F}\approx 0.15$ meV implies for the transverse dispersion at  $(k_x,k_y)=(0,0)$ a band-width of $2{(J_{\perp}^{\rm F})^2}/{\Delta}\approx 0.1$ meV, that can be measured with inelastic neutron
scattering. Finally, this model allows in principle to calculate the $T=0$ properties
at high field, i.e. beyond the field where classically the B-layers get populated. A proper treatment
of the vicinity of this field requires {\it a priori} to go beyond linear SW theory and is left for 
future investigation. It would be very useful however to measure the magnetization at 
very low temperature to see if a trace of this classical transition can be detected.

\acknowledgments
We gratefully acknowledge C. Berthier, M. Horvatic and C. R\"uegg for very useful discussions. This work has been supported by the Swiss National Fund, by MaNEP, and the r\'egion Midi-Pyr\'en\'ees through
its Chaire d'excellence Pierre de Fermat. FM also
thanks LPT (Toulouse) for hospitality.

\end{document}